\documentclass[prb,twocolumn,english,showpacs]{revtex4-1}
\usepackage{graphicx}
\usepackage{amssymb}

\begin{document}

\title{Observation of a pairing pseudogap in a two-dimensional Fermi gas}

\author{Michael Feld, Bernd Fr{\"o}hlich, Enrico Vogt, Marco Koschorreck, and Michael K{\"o}hl}

\affiliation{Cavendish Laboratory, University of Cambridge, JJ Thomson Avenue, Cambridge CB3 0HE, United Kingdom}

\date{\today}
\maketitle

\textbf{Pairing of fermions is ubiquitous in nature and it is responsible for a large variety of fascinating phenomena like superconductivity, superfluidity of $^3$He, the anomalous rotation of neutron stars, and the BEC-BCS crossover in strongly interacting Fermi gases. When confined to two dimensions, interacting many-body systems bear even more subtle effects\cite{Loktev2001}, many of which lack understanding at a fundamental level. Most striking is the, yet unexplained, effect of high-temperature superconductivity in cuprates, which is intimately related to the two-dimensional geometry of the crystal structure. In particular, the questions how many-body pairing is established at high temperature and whether it precedes superconductivity are crucial to be answered. Here, we report on the observation of pairing in a harmonically trapped two-dimensional atomic Fermi gas in the regime of strong coupling. We perform momentum-resolved photoemission spectroscopy\cite{Dao2007,Stewart2008}, analogous to ARPES in the solid state\cite{Damascelli2003}, to measure the spectral function of the gas and we detect a many-body pairing gap above the superfluid transition temperature. Our observations mark a significant step in the emulation of layered two-dimensional strongly correlated superconductors using ultracold atomic gases.}


One essential mechanism to establish collective quantum phases in Fermi systems, such as superconductivity or superfluidity, is fermionic pairing. In the case of weak attractive interactions between spin-up and spin-down fermions in three dimensions, Cooper pairs form on the surface of the Fermi sea, even though in vacuum no two-particle bound state exists. Pairing and the condensation of pairs go hand in hand. However, in this weak coupling limit the gap and hence the critical temperature $T_c$ are exponentially small. The fundamental and technological quest for increasing the superconducting transition temperature has led to the discovery of correlated materials which depart from weak coupling theory. These are complex compounds which combine strong interactions and large quantum fluctuations due to a two-dimensional geometry, both of which yet hinder a precise understanding and thorough theoretical modeling, even at zero temperature. Strongly correlated, two-dimensional, layered superconductors have displayed a wealth of new and unexpected phenomena. Among those is the peculiar observation of a suppression of low-energy weight in the single particle spectral function above the critical temperature and hence without direct link to superconducting behavior\cite{Ding1996}. The origin of this {\it pseudogap} phenomenon so far lacks full understanding and the implications of the pseudogap for the appearance of superconductivity have remained an open question.

Ultracold atomic gases provide an ideal testing ground to uncover the physics of strongly correlated Fermi systems because of the exceptional capability to tune the system parameters. Experiments in {\it three-dimensional} Fermi gases have observed fermionic many-body pairing in the crossover regime from BCS-like pairing to Bose-Einstein condensation (BEC) of dimers (for a review see \cite{Bloch2008}). Neither in the BCS- nor in the BEC-limit a pseudogap is expected in three dimensions. In between, i.e., for $k_F|a|\gg 1$, the gas is in the unitary regime of strongest interactions. The Fermi wave vector is $k_F$ and $a$ is the s-wave scattering length between spin-up and spin-down particles. Whether a pairing pseudogap phase exists here or whether a Fermi-liquid state dominates is still debated\cite{Tsuchiya2009,Haussmann2009,Gaebler2010,Hu2010,Chien2010,Nascimbene2011,Sommer2011,Wulin2011,Pieri2011}.

When the motion of particles is constrained to a two-dimensional plane, a bound state with binding energy $E_B$ exists for any value of the three-dimensional scattering length\cite{Petrov2001,Bloch2008,Frohlich2011}. This is in contrast to the situation in three dimensions where weakly bound molecules start to appear in the unitary regime of strongest interaction. The presence of the two-body bound state in vacuum is a necessary and sufficient condition for many-body pairing in a medium in two dimensions\cite{Randeria1989} and a small binding energy $E_B\ll E_F$ corresponds to the attractive BCS-like regime with interaction parameter $\ln(k_Fa_{2D})>1$. Here, $E_F$ denotes the Fermi energy, $a_{2D}=\hbar/\sqrt{mE_B}>0$ is the two-dimensional scattering length, $m$ is the mass of the atoms and $\hbar$ is Planck's constant divided by $2\pi$. At zero temperature, BCS-type superfluidity has been predicted in the attractive two-dimensional mean-field regime\cite{Miyake1983,Randeria1989,Drechsler1992,Bertaina2011}.

Qualitatively, we understand the behaviour at strong coupling and finite temperature by considering a complex order parameter\cite{Loktev2001} $\Delta(\textbf{x})=|\Delta(\textbf{x})|e^{i \theta(\textbf{x})}$. The phase $\theta(\textbf{x})$ may fluctuate as a function of the spatial coordinate $\textbf{x}$, for example driven by thermal fluctuations and quantum fluctuations enhanced in low-dimensional systems. For temperatures below the Berezinskii-Kosterlitz-Thouless transition temperature $T_{BKT}=T_c$, the superfluid gap $\Delta_{sc}=\langle \Delta\rangle$ is larger than zero. For higher temperature $T>T_{BKT}$ phase fluctuations destroy long range order (i.e. $\langle e^{i \theta(\textbf{x})}\rangle=0)$, however, the modulus $\Delta_{PG}=\langle|\Delta|\rangle$ remains finite up to the pairing crossover temperature $T^*$, which is on the order of $E_B$. The regime between $T_{BKT}$ and $T^*$ is referred to as the pseudogap regime and is still little understood. The transition between the normal phase and the pseudogap phase is not associated with breaking a continuous symmetry but is a crossover phenomenon. This scenario is different from standard weak-coupling mean-field BCS theory in which the superfluid gap is destroyed by pair breaking and $T_c$ and $T^*$ are approximately equal. The temperature range for accessing the pseudogap phase, i.e., the difference between $T^*$ and $T_c$, is particularly large in two dimensions since enhanced quantum fluctuations and second-order interaction effects\cite{Gorkov1961} suppress $T_{BKT}$ with respect to the mean-field result. The pseudogap regime is connected by a crossover to a Bose-liquid regime of local pairs when $\ln(k_Fa_{2D})$ crosses zero or, equivalently, $E_B=2E_F$.

\begin{figure}
 \includegraphics[width=\columnwidth,clip=true]{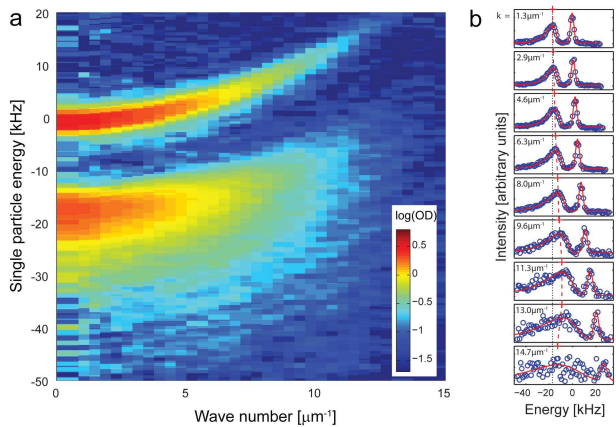}
 \caption{Measuring the spectral function. \textbf{a:} Measured photoemission signal at $\ln(k_Fa_{2D})=0$. The image is averaged over six repetitions and we plot $\log_{10}$ of the optical density.  \textbf{b:} Energy distribution curves for various values of the momentum $\hbar k$. The data of the pairing peak (lower feature) are fitted with a modified Gumbel distribution and the peak positions (red crosses) indicate the backbending of the dispersion curve. The data of the atomic peak are fitted with a Gaussian.}
 \label{fig1}
\end{figure}


Here, we study two-dimensional Fermi gases for $T>T_{BKT}$ in the strongly interacting regime $|\ln(k_Fa_{2D})|\leq 1$, in which the size of the pairs is comparable to the mean interparticle spacing. Hence, Pauli blocking and other many-body effects act on the pairing and a key question is, whether a pseudogap phase exists above the superfluid transition temperature\cite{Petrov2003,Bothelo2006} $T_{BKT} \approx 0.1 T_F$. We investigate many-body pairing using momentum-resolved photoemission spectroscopy, see methods. To this end, a long-wavelength photon of energy $\hbar \Omega$ creates a single particle excitation in the two-dimensional gas, which we detect at the momentum $k$. This measures the spectral function $A(\Omega,k)$ multiplied by the fermionic occupation function. This technique was originally developed for the study of solid state materials, such as cuprate superconductors\cite{Damascelli2003}, but recently has been successfully adapted to cold atomic systems\cite{Dao2007,Stewart2008}. Figure 1a shows a measured spectral function as a typical dataset used in this paper. The spectrum exhibits two features: The upper branch near zero energy corresponds to unpaired atoms, most likely from low-density regions of the trap with vanishingly small interaction between the free atoms\cite{Shin2007}. The lower branch corresponds to the pairing signal with a non-trivial dispersion in the spectral function. In figure 1b we plot the spectra $A(\Omega)$ for constant values of $k$, which show clearly asymmetric line shapes as expected for pair breaking into a continuum, together with our fits (see Methods).


\begin{figure*}
\includegraphics[width=\textwidth]{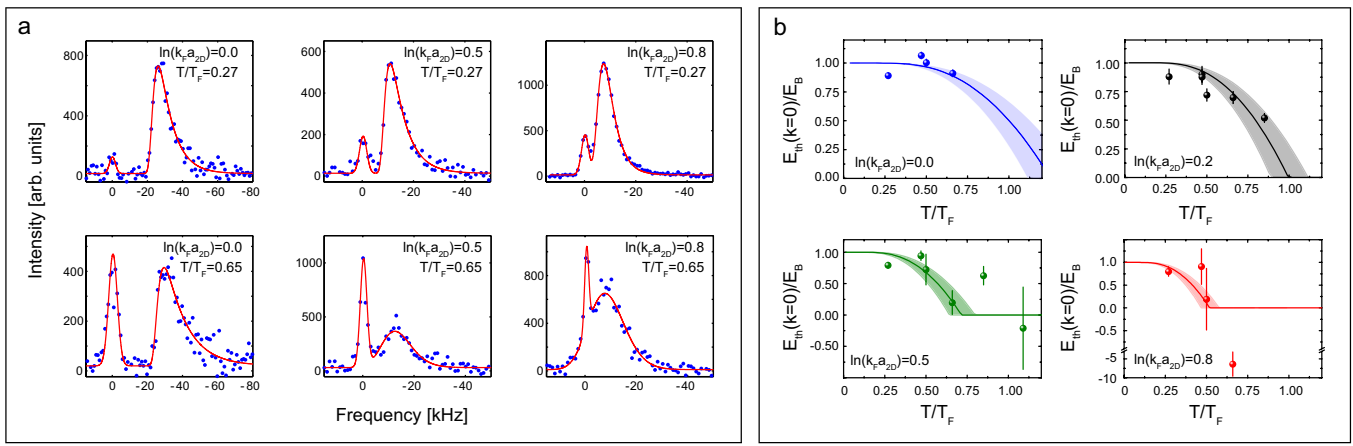}
 \caption{Pairing in the strongly interacting two-dimensional Fermi. \textbf{a}: Energy distribution curves $A(k=0,\Omega)$ for different values of $\ln(k_Fa_{2D})$ and temperatures. For low temperatures hardly any unpaired atoms exist. The solid line is the fit described in the text. \textbf{b}: Energy threshold of the pairing peak as a function of temperature for different interaction strengths. The solid lines are predictions of mean-field theory\cite{Loktev2001} for the quantity $|\Delta(T)|^2/2E_F$ in two dimensions without free parameters. The shaded area marks the uncertainty of our determination of $\ln(k_Fa_{2D})$, which is $\pm0.1$. The error bars show the 1-$\sigma$ fit errors. They become larger towards larger values of $\ln(k_Fa_{2D})$ because of the division by the exponentially smaller binding energy $E_B$. }
 \label{fig2}
\end{figure*}

We analyze the spectral function for various interaction strengths and temperatures. In the attractively interacting regime $\ln(k_Fa_{2D}) >0$ the size $a_{2D}$ of the pairs is larger than the mean interparticle spacing $k_F^{-1}$. Therefore, at low temperature no isolated dimers exist in the sample but a many-body pairing gap opens in the spectrum, conceptually similar to a BCS-paired state in three dimensions. The zero-temperature limit of the two-dimensional BEC-BCS crossover has been investigated in the framework of mean-field theory\cite{Randeria1989,Loktev2001}. This has led to the remarkable prediction that at zero temperature the condensation energy per particle $E_0=\Delta_{sc}^2/2E_F$ at $k=0$ should be equal to the two-body binding energy in vacuum $E_B$ for all interaction strengths.

The expected rf spectrum can be qualitatively estimated based on mean-field theory of a BCS state at zero temperature. Many-body pairing causes an energy threshold $E_{th}$ and this energy has to be supplied by the rf field to create a single-particle excitation. Ignoring final state interactions and assuming two-dimensional geometry with a constant density of states, the momentum integrated lineshape has the form $\Gamma(\Omega)\propto \Theta(\hbar \Omega-E_{th})/\Omega^2$. Here, $\Theta(x)$ is the Heaviside function. We investigate many-body pairing at low temperatures by studying the energy threshold $E_{th}$ of the pairing peak in the photoemission spectrum at momentum $k=0$. This quantity is closely linked to $E_0$ and its disappearance, for example as temperature is increased, indicates the transition into a normal phase. Investigating $E_{th}$ at $k=0$, rather than for example at $k_F$, has the advantage that averaging over the inhomogeneous spread of $k_F$ and $\mu$ of the approximately 30 two-dimensional gases has a small effect because the mean-field prediction at zero temperature is not explicitly dependent on $\mu$ and $k_F$. At the temperature $T/T_F=0.27$ (see Figure 2a, top row), we observe a negligible contribution of free atoms and a sharp onset of the pairing peak. As we increase the temperature of our sample (see Figure 2a, bottom row) the spectral weight at low energies increases because the pairing peak broadens and the threshold shifts. For $\ln(k_Fa_{2D})=0.8$ at $T/T_F=0.65$, for example, the broadening creates a significant amount of low-energy spectral weight and the threshold of the spectrum even moves to positive energies. The latter result shows that the energy gap has completely disappeared.

In Figure 2b we show the measured temperature dependence of $E_{th}$ for various interaction strengths on the BCS side of the resonance $\ln(k_Fa_{2D})>0$, where no local dimers exist in the sample. We observe that $E_{th}$ moves towards zero with increasing temperature, which indicates that the many-body pairing gap vanishes. We compare our data to finite temperature mean-field theory in two dimensions\cite{Loktev2001}. Overall, the numerically calculated energy $E_0=|\Delta(T)|^2/2E_F$ reproduces our experimental data for $E_{th}$ without free parameters (solid line in Figure 2b). However, some discrepancies remain. For example, for the coldest temperatures the average of $E_{th}/E_B$ over all interaction strengths is $E_{th}/E_B=(0.84\pm0.05)$, approximately $15\%$ smaller than predicted by mean-field theory. The deviation could stem from beyond mean-field effects provoked by our two-dimensional geometry and interaction energy shifts. It could be more precisely computed using Quantum-Monte Carlo calculations, which, however, up to now exist only for zero temperature\cite{Bertaina2011}. From our data we conclude that we have realized a many-body pairing phase in two dimensions and observed how it vanishes as the temperature is increased to approach the pairing crossover temperature $T^*$. Our data are taken above the critical temperature for superfluidity $T_{BKT}$ in two dimensions, which is predicted to be at $T/T_F \approx 0.1$ in the strong coupling regime. We have verified this experimentally by employing a rapid magnetic field sweep across the Feshbach resonance to project the possibly condensed pairs onto deeply bound molecules and then image the molecular condensate after time-of-flight\cite{Regal2004a}. While this procedure allows us to observe three-dimensional Fermi condensates, in our two-dimensional samples no condensation was detected. We therefore identify our phase as a pairing pseudogap phase.

\begin{figure}
 \includegraphics[width=\columnwidth,clip=true]{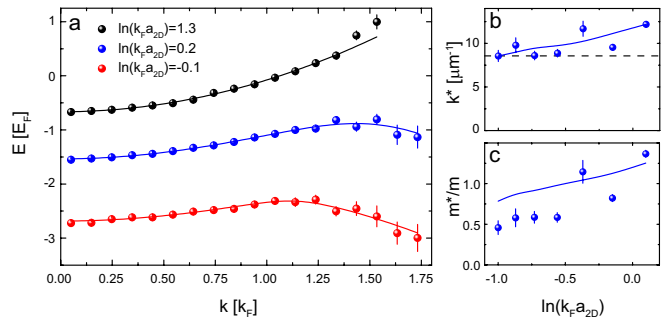}
 \caption{Interaction dependence of the quasiparticle dispersion. \textbf{a:} Dispersion for different interaction strengths at $T/T_F=0.27$. In the strongly interacting crossover regime, we observe back-bending of the dispersion relation which vanishes in the regime of small interaction. The solid line is the fit of a BCS-like dispersion relation defined in the text. \textbf{b:} Wave vector $k^*$ at which the back-bending of the spectral function occurs, extracted from a fit to the dispersion relation. The dashed line indicates the Fermi wave vector. \textbf{c:} Effective mass, extracted from a fit to the dispersion relation. The solid lines in b and c are the prediction of a thermal singlet model. The error bars reflect the fitting errors.}
 \label{fig3}
\end{figure}

In addition to the energy threshold at zero momentum, we also investigate how the quasiparticle dispersion is affected when changing temperature and interaction strength. We extract the dispersion data from the peak of the spectral function $E_{peak}(k)=\max_\Omega\left[A(k,\Omega)\right]$ (see Figure 1b). In the crossover region and on the BEC side of the resonance, the dispersion shows a pronounced back-bending, which we display in Figure 3a for various interaction strengths. This shape is reminiscent of the hole dispersion of a BCS state and qualitatively similar to that of a three-dimensional Fermi gas near unitarity\cite{Stewart2008}. We fit the dispersion relation $E(k)=\mu -\sqrt{\left[\hbar^2/(2 m^*)(k^2-k^{*2})\right]^2+\Delta^2}$ for $k\leq 1.5 k_F$ using the zero temperature mean-field prediction for the chemical potential\cite{Randeria1989} to extract the parameters effective mass $m^*$ and the wave vector $k^*$, at which the backbending occurs. In Figures 3b and 3c we display the results of $k^*$ and $m^*$, respectively, which show that $k^*$ remains similar to $k_F=8.1\,\mu$m$^{-1}$ across the whole crossover. We note that the temperature of the cloud increases on the far BEC side, possibly due to heating introduced by the nearby confinement-induced p-wave resonance\cite{Gunter2005}. We model our data on the BEC side of the resonance with a thermal ensemble of singlet pairs using the exact expression of the pair wave function in two dimensions and the experimentally determined temperatures \cite{Gaebler2010} (see Methods). The approximation of phase-disordered pairs in the normal state includes the correct short-range physics but neglects interactions between pairs as well as Pauli-blocking in the dissociation process. Our numerical data show that for experimentally relevant parameters the thermal singlet model can give rise to a back-bending feature of the photoemission spectrum with similar values of $m^*$ and $k^*$ as in the experiment (see solid lines in Figure 3b and c). A contribution to the back-bending effect from incoherent pairs has been predicted previously in three dimensions\cite{Schneider2010b}. We note that, despite working reasonably well for $k^*$ and $m^*$, the thermal singlet model makes incorrect predictions for the energy threshold in Figure 2.


It is instructive to observe what happens to the dispersion signal at elevated temperatures. To this end, we have performed experiments in the range $0.27\leq T/T_F \leq 0.65$ in the strongly interacting crossover regime. We observe that the BCS-like dispersion relation converts into a free-particle like dispersion as the temperature is increased (see Figure 4a-c) and we determine the temperature $T_B^*$ at which this change takes place. Figure 4d shows the variation of $T_B^*$ (in units of $E_B$) in the crossover regime and on the BEC side. We find that the temperature $T_B^*$ is comparable to the pairing temperature $T^*$ of mean-field theory (dashed line in Figure 4d), however, a factor of $0.36$ smaller. This discrepancy could suggest that the appearance of the back-bending feature in the spectral function, which has been interpreted as a signature for many-body pairing\cite{Stewart2008,Gaebler2010}, is mainly a qualitative evidence. In contrast, our spectroscopic measurements of the energy threshold (see Figure 2) give quantitative results for many-body pairing and agreement with theory even in a regime where no back-bending feature is observable.

\begin{figure}
\includegraphics[width=\columnwidth]{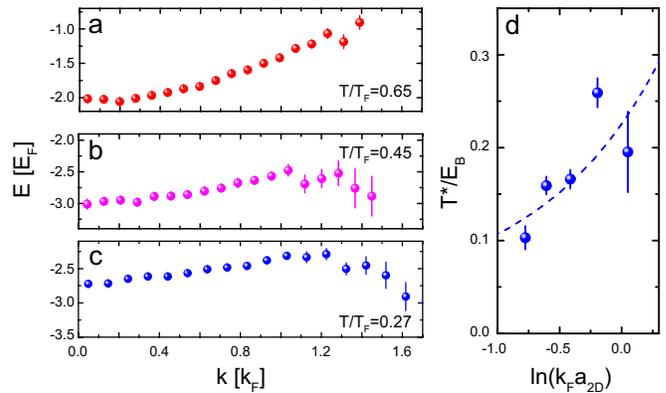}
 \caption{Temperature dependence of the quasiparticle dispersion. \textbf{a-c:}Temperature dependence of the backbending signature of the quasiparticle dispersion at $\ln(k_Fa_{2D})=-0.2$. \textbf{d:} Evolution of the pairing temperature $T_B^*$ as a function of the interaction parameter $\ln(k_F a_{2D})$. The dashed line is the theoretical prediction of $T^*$ from mean-field theory, scaled by a factor of 0.36.}
 \label{fig4}
\end{figure}

In conclusion, we have observed a many-body pairing gap above the superfluid transition temperature of a strongly interacting two-dimensional Fermi gas. Our results represent the first major step in emulating and understanding pairing in two-dimensional, strongly correlated materials using cold atoms. In future experiments, an even closer cross-link to cuprate high-temperature superconductors could be provided by the inclusion of an optical lattice potential to realize the two-dimensional Fermi-Hubbard model. In this model, a pseudogap phase is expected for a slightly doped antiferromagnetic state and it is believed to precede exotic quantum states like the d-wave superconducting phase, which will be detectable by our technique.

\section*{Methods summary}

In our experimental setup\cite{Frohlich2011}, we prepare a quantum degenerate Fermi gas of $^{40}$K atoms in a 50/50 mixture of the two lowest hyperfine states $|F=9/2,m_F=-9/2\rangle$ and $|F=9/2,m_F=-7/2\rangle$. We confine the quantum gas to two dimensions in a deep optical lattice formed by a standing wave laser field, preparing approximately 30 layers. The interaction strength between spin-up and spin-down particles is tuned at a Feshbach resonance near 202.1\,Gauss. The photoemission measurement couples the $|F=9/2,m_F=-7/2\rangle$ state to the weakly interacting state $|F=9/2,m_F=-5/2\rangle$ using a radiofrequency photon of frequency $\Omega$ with negligible momentum transfer. We measure the momentum distribution of the transferred atoms in a time-of-flight experiment and average the absorption signal azimuthally to obtain $A(k,\Omega)$, where $k=\sqrt{k_x^2+k_y^2}$.

\section*{methods}

\subsection{Experimental setup}
We evaporatively cool a 50/50 spin mixture of $^{40}$K atoms in the $|F=9/2,m_F=-9/2\rangle\equiv |-9/2\rangle$ and $|F=9/2,m_F=-7/2\rangle\equiv |-7/2\rangle$ states of the hyperfine ground state manifold\cite{Frohlich2011}. After reaching quantum degeneracy in a crossed-beam optical dipole trap with approximately 70000 atoms per spin state, we turn on an optical lattice potential in order to prepare two-dimensional Fermi gases\cite{Gunter2005,Martiyanov2010,Frohlich2011,Dyke2011}. The optical lattice is formed by a horizontally propagating, retro-reflected laser beam of wavelength $\lambda=1064$\,nm, focussed to a waist of 140\,$\mu$m. We increase the laser power over a time of 200\,ms to reach a final potential depth of up to $V_{lat}=83\,E_{rec}$, which is calibrated by intensity modulation spectroscopy. $E_{rec}=h^2/(2 m \lambda^2)$ is the recoil energy. The trapping frequency along the strongly confined direction is $\omega=2 \pi \times 78.5$\,kHz. After loading the optical lattice, we adiabatically reduce the power of the optical dipole trap such that the atoms are confined only by the Gaussian intensity envelope of the lattice laser beams. The radial trapping frequency of the two-dimensional gases is $\omega_\perp=2\pi\times 127$\,Hz for $V_{lat}=83\,E_{rec}$ and we confine on the order of $10^3$ atoms per two-dimensional gas at the center of the trap. Along the axial direction we populate approximately 30 layers of the optical lattice potential with an inhomogeneous peak density distribution. Approximately two thirds of the 2D layers with highest density dominate the measured signal and their relevant energy scales $E_F$, $E_B$, and $\Delta^2/2E_F$ are more than an order of magnitude larger than the trapping frequency $\omega_\perp$. Therefore, finite particle number effects do not influence the measured signal. After evaporation, we adiabatically increase the interaction strength by lowering the magnetic field, at a rate of up to 0.25\,G/ms, to a value near the Feshbach resonance at 202.1\,G. We apply a radio-frequency pulse near 47\,MHz with a Gaussian amplitude envelope with a full width at half maximum of 230\,$\mu$s to transfer atoms from the $|-7/2\rangle$ state to the $|F=9/2,m_F=-5/2\rangle$ state. Atoms in the $|9/2,-5/2\rangle$ state have a two-body s-wave scattering length of 130 Bohr radii with the $|-7/2\rangle$ state and 250 Bohr radii with the $|-9/2\rangle$ state\cite{Stewart2008}. We turn off the optical lattice 100\,$\mu$s after the radiofrequency pulse, switch off the magnetic field and apply a magnetic field gradient to achieve spatial splitting of the three spin components in a Stern-Gerlach experiment. For each run, the magnetic field is calibrated using spin-rotation with an rf pulse of an imbalanced mixture on the $|-9/2\rangle$/$|-7/2\rangle$ transition. The magnetic field accuracy deduced from these measurements is $<3$\,mG. We measure the temperature by ballistic expansion of a weakly interacting gas, and the quoted numbers refer to the average of $T/T_F$ across the whole sample.

\subsection{Determination of the energy threshold $E_{th}$ of the energy distribution curve}
We fit our data with a double-peak fitting function comprising of a Gaussian for the atomic signal and a modified Gumbel function $f(\Omega)=\alpha \exp[-(\Omega-\Omega_0)/b-a \exp(-(\Omega-\Omega_0)/(a b))]$ for the pairing peak. The parameter $\Omega_0$ measures the peak position and the parameters $a$ and $b$ measure skewness and width. For our further analysis, we only use the peak position $\Omega_0$, which does not depend on the line shape function used. From this fit we determine the maximum of the molecular peak $\nu_{max}=\Omega_0$ and the minimum between the atomic and the molecular peak $\nu_{min}$. Between $\nu_1=\nu_{max}$ and $\nu_2=\nu_{min}-2$\,kHz we fit the data with a linear function and determine the zero-crossing of the linear extrapolation as the energy threshold $E_{th}$. We correct the obtained result for our spectral resolution of $1.5$\,kHz, obtained from the width of the Gaussian fits. The data are normalized to the two-body binding energy in vacuum which we obtain from the transcendental equation\cite{Bloch2008}
\begin{equation}
l_z/a=\int_0^\infty \frac{du}{\sqrt{4 \pi u^3}} \left(1- \frac{\exp(-E_B u/(\hbar \omega))}{\sqrt{(1-\exp(-2 u))/(2 u)}}\right).
\end{equation}
Here, $l_z=\sqrt{\hbar/m\omega}$ and $a$ is the three-dimensional scattering length using the following parameters of the Feshbach resonance: $B_0=202.1$\,G, $\Delta B=7$\,G and $a_{BG}=174\,a_B$ where $a_B$ is the Bohr radius.

\subsection{Thermal singlet model}
We model our data on the BEC side of the resonance with a thermal ensemble of singlet pairs\cite{Gaebler2010}. The expression for the wave function of the bound state in two dimensions is $\psi_B(r)=\sqrt{2/a_{2D}}K_0(r/a_{2D})$, in which $K_0(x)$ is the modified Bessel function, and for the scattering state is $\psi_q(r) = J_0(qr) -\frac{i f(q)}{4} H^{(1)}_0 (qr)$, in which $J_0(x)$ is the Bessel function of the first kind and $H^{(1)}_0(x)$ is the Hankel function of the first kind\cite{Petrov2001}. $f(q)$ is the scattering amplitude between the state $|-7/2\rangle$ and the final state $|-5/2\rangle$. We compute the momentum resolved rf spectrum for the dissociation from the bound state to the scattering state, averaging over a thermal distribution of the center-of-mass momenta of the initial pairs using  Monte-Carlo sampling.  From the momentum-resolved rf spectrum we calculate the effective mass $m^*$ and the wave vector $k^*$ using the same fitting routines as for the experimental data. This model of tightly bound pairs in the normal state includes the correct short-range physics but neglects many-body pairing, interactions between atoms and between pairs, as well as quantum statistical effects. Therefore, we do not expect quantitative agreement in the strongly interacting regime or on the BCS side of the resonance.

\vspace{0.5 cm}

 We thank A. Georges, C. Kollath, D. Pertot, D. Petrov, M. Randeria, W. Zwerger, and M. Zwierlein for discussions. The work has been supported by {EPSRC} (EP/G029547/1), Daimler-Benz Foundation (B.F.), Studienstiftung, and DAAD (M.F.).

 The authors declare that they have no competing financial interests.

 The experimental setup was devised and constructed by M.F., B.F., E.V., and M.K., data taking was performed by M.F., B.F., E.V., and M. Kos., data analysis was performed by M.F., B.F., and M.Kos., numerical modelling was performed by B.F., and the manuscript was written by M.K. with contributions from all coauthors.

 Correspondence and requests for materials should be addressed to M.K.~(email: mk540@cam.ac.uk).

\end{document}